\documentclass[t12pt,pra]{revtex4}
\usepackage{graphicx}
\usepackage{epsfig}
\usepackage{latexsym}
\usepackage{amssymb}
\usepackage{bm}
\usepackage{float}
\usepackage{subfigure}
\usepackage{amsmath}
\usepackage{amsfonts}

\def\T{{\cal T} }
\def\m{{M_{\text{Planck}}} }

\begin{document}

\begin{flushright}
\small{~~~~\\ ~~~~~~~~~~~~~~~~~~~~~~~~~~~~~~~~~~~~~~~~~~~~~~~~~~~~~~~~~~~~~~~ UTTG-22-16 }\
\end{flushright}

\title{Physical Effects of the Gravitational $\Theta$-Parameter\footnote{This essay received on Honorable Mention in the 2016 Essay Competition of the Gravity Research
Foundation.}}

\author{Willy Fischler}
\email{fischler@physics.utexas.edu}
\affiliation{Theory Group, Department of Physics, University of Texas, Austin, TX 78712, USA \\ and \\ Texas Cosmology Center, University of Texas, Austin, TX 78712, USA}
\author{Sandipan Kundu}
\email{kundu@cornell.edu}
\affiliation{Department of Physics, Cornell University, Ithaca, New York, 14853, USA}

\begin{abstract}

We describe the effect of the gravitational $\Theta$-parameter on the behavior of the stretched horizon of a black hole in $(3+1)$-dimensions. The gravitational $\Theta$-term is a total derivative, however, it affects the transport properties of the horizon. In particular,  the horizon acquires a third order parity violating, dimensionless transport coefficient which affects the way localized perturbations scramble on the horizon. In the context of the gauge/gravity duality, the $\Theta$-term induces a non-trivial contact term in the energy-momentum tensor of a $(2+1)-$dimensional large-N gauge theory, which admits a dual gravity description. As a consequence, the dual gauge theory in the presence of the $\Theta$-term acquires the same third order parity violating transport coefficient.
\end{abstract}

\maketitle

In a sensible theory of quantum gravity, the leading term in the low energy limit should be the Einstein-Hilbert action, however, it is expected that the low energy effective theory will also contain higher derivative correction terms. In order $R^2$, there is only one ghost-free combination, known as the Gauss-Bonnet term. In $(3+1)-$dimensions, the Gauss-Bonnet term is a total derivative, however, in the presence of this topological term the entropy of a black hole receives a correction which is proportional to the Euler number of the horizon. It has been established that this correction term violates the second law of black hole thermodynamics and hence the Gauss-Bonnet term  is problematic \cite{Jacobson:1993xs,Olea:2005gb, Liko:2007vi, Sarkar:2010xp}.

In particle physics, it is well known that total derivative terms can have physical effects. Lorentz and gauge invariance of quantum chromodynamics (QCD) allow for a CP-violating topological $\theta$-term which contributes to the electric dipole moment of  neutrons \cite{Crewther:1979pi}. In $(3+1)-$dimensions, similar to the QCD-$\theta$ term, there exists a parity violating gravitational $\Theta$-term \cite{theta}
\begin{equation}\label{theta}
 S_\Theta=\frac{\Theta}{8}\int d^4x\ \epsilon^{\mu \nu \alpha \beta}R^\tau_{~\sigma \mu \nu} R^{\sigma}_{~\tau \alpha \beta}\ .
\end{equation}
The gravitational $\Theta$-term is an elusive quantity; it does not contribute to the classical equations of motion because it is a total derivative. However, by using the membrane paradigm, we will show that this $\Theta$-parameter does affect the way localized perturbations scramble on black hole horizons.

The membrane paradigm  \cite{Price:1986yy,Thorne:1986iy} enables us to study the interactions of black holes with the outside by replacing the true mathematical horizon by a {\it stretched horizon}, an effective time-like membrane located roughly one Planck length away from the true horizon. The black hole entropy is finite which suggests that the effective number of degrees of freedom between the actual black hole horizon and the stretched horizon  should be negligibly small. Hence, it is natural as well as convenient to replace the true horizon by a stretched horizon. Predictions of the membrane paradigm are considered to be robust since they depend on some very general assumptions: 

\begin{itemize}
\item{ Between the actual black hole horizon and the stretched horizon, the effective number of degrees of freedom are vanishingly small.} 
\item { Physics outside the black hole, classically must not be affected by the dynamics inside the black hole. }
\end{itemize}

One of the reasons that black holes are fascinating, is that they do provide us with a natural laboratory to perform thought experiments that shed some light on quantum gravity. String theory, Matrix Theory \cite{Banks:1996vh}, and the AdS/CFT correspondence \cite{Maldacena:1997re} are the only models of quantum gravity over which we have mathematical control. These models have provided us with some insight into different aspects of quantum gravity, however, none of them give us a complete microscopic description of the physics of black holes. Historically the membrane paradigm has also been successful at providing us with an elegant and powerful framework to study macroscopic properties of black hole horizons. In particular,  the membrane paradigm has been used extensively as an efficient computational tool to study various astrophysical phenomena in the vicinity of black holes. In addition, the membrane paradigm has also been able to provide crucial hints about the nature of the microscopic physics of horizons. For example, it predicts that black hole horizons are the {\it fastest scramblers} of information in nature which strongly indicates that the microscopic description of black hole horizons must involve non-local degrees of freedom \cite{Sekino:2008he}.

The membrane paradigm tells us that for an outside observer a black hole horizon effectively behaves like a viscous Newtonian fluid for the Einstein gravity \cite{Price:1986yy,Thorne:1986iy}. In the presence of the gravitational $\Theta$-term (\ref{theta}), the three-dimensional stretched horizon energy-momentum tensor receives a correction \cite{Fischler:2015kro}
\begin{equation}\label{theta_em}
 \T_{i j}=\left[\m^2\left(h_{i j}K -K_{ij}\right) - 2 \Theta C_{ij}\right]_{\text Stretched\ Horizon}\ ,
\end{equation}
where  $C_{mn}=-\frac{e^{ijk}}{2}[h_{mk} D^{(3)}_{i} R^{(3)}_{nj}+h_{nk} D^{(3)}_{i} R^{(3)}_{mj}]$ is the Cotton-York tensor which is symmetric, traceless and covariantly conserved in three-dimensions. $h_{ij}$, $K_{ij}$ and $R^{(3)}_{ij}$ are  respectively induced metric, extrinsic curvature and Ricci tensor of the stretched horizon. Terms proportional to $\m^2$ in (\ref{theta_em}) come from the Einstein gravity and they are responsible for the viscous fluid like behavior of black hole horizons. Different first order transport coefficients of this fluid for the Einstein gravity were first derived in \cite{Price:1986yy}. The gravitational $\Theta$-term contribution of (\ref{theta_em}) affects the transport properties of the horizon fluid, in particular,  the horizon fluid acquires a third order parity violating, dimensionless transport coefficient, which we will call $\vartheta$.

This is a new hydrodynamic effect, so let us briefly describe some of the properties of $\vartheta$. It is a parity violating third order  transport coefficient  in $(2+1)-$dimensions and hence forbidden in a parity-invariant theory. Transport coefficients measure the response of a fluid after a hydrodynamic perturbation. For example, one can disturb a hydrodynamic system by perturbing the background metric and then observe the evolution of the energy momentum tensor of the fluid as a result of the metric perturbation. Under a small metric perturbation $\gamma_{AB}$ around flat Minkwoski metric, $\vartheta$ contributes to the energy-momentum tensor in the following way:
\begin{align}\label{third1}
T_{11}=-T_{22}=-\vartheta \frac{\partial^3 \gamma_{12}}{\partial t^3}\ , \qquad
T_{12}=T_{21}=\frac{\vartheta}{2 }\left( \frac{\partial^3 \gamma_{11}}{\partial t^3}- \frac{\partial^3 \gamma_{22}}{\partial t^3}\right) \ 
\end{align}
and hence $\vartheta$ contributes to the retarded Green's function of the energy-momentum tensor in order $\omega^3$:
\begin{align}
G^R_{12,11-22}(\omega,\vec{k}\rightarrow 0)=-2i \vartheta \omega^3\ .
\end{align}
$\vartheta$ is dimensionless and it does not affect the trace of the energy-momentum tensor. In $(2+1)-$dimensional hydrodynamics, the {\it Hall viscosity} is another parity violating transport coefficient that appears in the first order in derivative expansion. The Hall viscosity has been studied extensively for both relativistic and non-relativistic systems. $\vartheta$ is a third order cousin of Hall viscosity and hence it is also an example of Berry-like transport \cite{Haehl:2014zda}. In the presence of the $\Theta$-term, the horizon acquires this ``Hall viscosity-like" transport coefficient $\vartheta$, with $\vartheta=\Theta$ \cite{Fischler:2015kro}. It is truly amazing that we can discover a new hydrodynamic effect by studying fluctuations on the stretched horizon in the presence of the $\Theta$-term.

Physically, the appearance of this new transport coefficient indicates that the $\Theta$-term affects the way perturbations scramble on the horizon. The process by which a localized perturbation spreads out into the whole system is known as {\it scrambling}. In quantum mechanics, information contained  inside a small subsystem of a larger system  is said to be fully scrambled when the small subsystem becomes entangled with the rest of the system. After scrambling time $t_s$  the information can only be retrieved by examining practically all the degrees of freedom. In a local quantum field theory, the scrambling time $t_s$ is expected to be at least as long as the diffusion time. Consequently, one can show that the scrambling time for a strongly correlated quantum fluid in $d$-spatial dimensions satisfies $t_s T \ge c \hbar S^{2/d}$, where $c$ is some dimensionless constant, $T$ is the temperature  and $S$ is the total entropy. It has been argued in \cite{Sekino:2008he} that this has to be a universal bound on the scrambling time. Hence, it is indeed remarkable, as shown in  \cite{Sekino:2008he}, that black hole horizons scramble information exponentially fast $t_s T \approx \hbar \ln S$, violating the bound. This unusual process is known as ``fast-scrambling" and it strongly suggests that the microscopic description of fast-scrambling on black hole horizons must involve non-local degrees of freedom \cite{Sekino:2008he}. In the presence of the gravitational $\Theta$-term, let us now perform a thought experiment in which an outside observer drops a massive particle onto the black hole and sees how the perturbation scrambles on the horizon. Equation (\ref{third1}) indicate that the gravitational $\Theta$-term will  affect the way perturbations scramble on the horizon, in particular, similar to the electrodynamics $\theta$-angle  \cite{Fischler:2015cma}, it will also introduce vortices without changing the scrambling time \cite{Fischler:2015kro}. This strongly suggests that in a sensible theory of quantum gravity the $\Theta$-term will play an important role, a claim that is also supported by the AdS/CFT correspondence.

\bigskip

The membrane paradigm has become even more relevant with the emergence of {\it holography} \cite{Banks:1996vh, Maldacena:1997re}, a remarkable idea that connects two cornerstones of theoretical physics: quantum gravity and gauge theory. The AdS/CFT correspondence \cite{Maldacena:1997re}, which is a concrete realization of this idea of holography, has successfully provided us with theoretical control over a large class of strongly interacting field theories. This duality enables us to compute observables of certain gauge theories in $d$-dimensions by performing some classical gravity calculations in $(d+1)$-dimensions. Gravity duals of these field theories at finite temperature contain black holes, where the field theory temperature is given by the Hawking temperature of black holes.  It is well known that there is a connection between the low frequency limit of the linear response of a strongly coupled quantum field theory and the membrane paradigm fluid on the black hole horizon of the dual gravity theory \cite{Kovtun:2003wp,Kovtun:2004de,Son:2007vk,Iqbal:2008by,Bredberg:2010ky}. Long before the emergence of gauge/gravity duality, it was shown that the membrane paradigm fluid on the stretched horizon of a black hole has a shear viscosity to entropy density ratio of $1/4\pi$ \cite{Price:1986yy, Thorne:1986iy}. Later it was found that the shear viscosity to entropy density ratio of a gauge theory with a gravity dual is indeed $1/4\pi$ \cite{Kovtun:2004de,Bredberg:2010ky}. Interestingly, the strong coupling physics of quark-gluon plasma has been experimentally explored in the Relativistic Heavy Ion Collider (RHIC), where the shear viscosity to entropy density ratio has been measured to be close to $1/4\pi$. 

Let us now consider a large-N gauge theory in $(2+1)-$dimensions which is dual to a gravity theory in $(3+1)-$dimensions with the gravitational $\Theta$-term and discuss the effect of the parity violating $\Theta$-term on the dual field theory. A reasonable guess is that the dual gauge theory, similar to the membrane paradigm fluid, will acquire the same third order parity violating transport coefficient $\vartheta$.  This guess has been confirmed by explicit computation in \cite{Fischler:2015kro}.

It was first pointed out in \cite{Closset:2012vp} that the two-point function of the energy-momentum tensor in a $(2+1)-$dimensional conformal field theory can have a non-trivial contact term
\begin{align}\label{cft}
\langle T_{ij}(x)T_{mn}(0)\rangle=-i\frac{\kappa_g}{192\pi} \left[\left(\varepsilon_{iml}\partial^l\left(\partial_j \partial_n-\partial^2 \delta_{jn} \right) +(i\leftrightarrow j)\right)+(m\leftrightarrow n)\right]\delta^3(x)\ .
\end{align}
It is possible to shift $\kappa_g$ by an integer by adding a gravitational Chern-Simons counterterm to the UV-Lagrangian and hence the integer part of $\kappa_g$ is scheme-dependent. On the other hand, the fractional part $\kappa_g$ mod $1$  does not depend on short distance physics and hence is a meaningful physical observable in a $(2+1)-$dimensional conformal field theory \cite{Closset:2012vp}. Following the AdS/CFT dictionary, we can easily calculate the two-point function of energy-momentum tensor for a large-N conformal field theory in $(2+1)-$dimensions which is dual to an anti-de-Sitter spacetime in $(3+1)-$dimension  with the gravitational $\Theta$-term. The $\Theta$-term induces a conformally invariant contact term in the two-point function of the energy-momentum tensor of the dual field theory which has the form (\ref{cft}) with $\kappa_g=96\pi\Theta$ \cite{Fischler:2015kro}. Therefore, a gravity theory in AdS$_{3+1}$  with the gravitational $\Theta$-term is dual to a conformal field theory with  non-vanishing $\kappa_g$. This also suggests that only a fractional part of the $\Theta$-term is a well-defined observable.

The contact term $\kappa_g$ is also related to the transport coefficient $\vartheta$, in particular for a holographic theory dual to asymptotically AdS spacetime in $(3+1)-$dimensions: $\vartheta=\Theta=\kappa_g/96\pi$. Note that for a holographic theory, $\vartheta$ is independent of the temperature. This is a nice example where gravity teaches us about a new hydrodynamic effect and suggests an interesting study of the transport coefficient $\vartheta$ for condensed matter systems.  Historically, gauge/gravity duality has played an important role in hydrodynamics by discovering new universal effects \cite{Bhattacharyya:2008jc,Erdmenger:2008rm,Son:2009tf}.  However, it is important to stress that one could have found this hydrodynamic effect without knowing anything about the AdS/CFT correspondence by simply studying the effect of the $\Theta$-term on the stretched horizon.

\bigskip

It is important to note that our discussion about the effect of the $\Theta$-term on the stretched horizon can easily be generalized to arbitrary cosmological horizons. The AdS/CFT correspondence has taught us that the membrane paradigm fluid on the black hole horizon and the linear response of a strongly coupled quantum field theory in the low frequency limit, are related. However, it is not at all clear if this connection between the membrane paradigm and holography goes beyond the AdS/CFT correspondence. In particular, it will be extremely interesting to understand if the same conclusion holds for holographic models of cosmological spacetime.

\end{document}